\documentclass[referee]{raa}            

\usepackage{graphicx,times}             
\usepackage{natbib}
\usepackage{amssymb,amsmath}
\bibpunct{(}{)}{;}{a}{}{,}

\usepackage[driverfallback=dvipdfm]{hyperref}

\hypersetup{colorlinks = true, linkcolor = red, anchorcolor = red, citecolor = blue, filecolor = red, pagecolor = red, urlcolor = red}

\begin{document}

   \title{Physical parameters of W UMa type contact binaries and its stability of mass transfer}

   \volnopage{Vol.0 (20xx) No.0, 000--000}      
   \setcounter{page}{1}          

   \author{Berikol Tekeste Gebreyesus\inst{1,2}, Seblu Humne Negu\inst{1}}

   \institute{\inst{1} Space Science and Giospacial Institute (SSGI),
              Entoto Observatory and Research Center (EORC)
              Astronomy and Astrophysics Department,
              P.~O.~Box 33679 Addis Ababa, Ethiopia. {\it  berikoltekeste@gmail.com}, {\it seblu1557@gmail.com}\\
              \inst{2} Addis Ababa University, P.~O.~Box 1176 Addis Ababa Ethiopia. {\it  berikoltekeste@gmail.com} \\
\vs\no
   {\small Received~~20xx month day; accepted~~20xx~~month day}}

\abstract {In this study, we determined the physical parameters of W UMa type contact binaries and its stability of mass transfer with 
different stellar mass ranges over a broad space
by applying the basic dynamical evolution equations of the W UMa type contact binaries using 
accretor and donor masses between 0.079 and 2.79 $M_{\odot}$.
In these systems, we have studied the three sub-classes of W UMa system such as A-, B- 
and W-type of W UMa contact binaries using the initial and final mass ranges and we 
investigated different stellar and orbital parameters for the sub-classes of W UMa systems. 
We examined the stability of the W UMa type contact binaries using the orbital parameters such as 
critical mass ratio, Roche lobe radius of the donor star and mass ratio of these systems.
Thus, we computed the observed and calculated physical parameters of A-, B- and W-type of W UMa systems.
Although, we determined the combined and color temperatures to classify the three subclass of the systems.
Also, we have presented the result of the internal stellar structure and evolution of W UMa type contact binaries by using the polytropic model.
}

   \authorrunning{B.T. Gebreyesus and S. H. Negu}            
   \titlerunning{Physical parameters of W UMa type CBs }  
   \maketitle

%
\newpage
\section{Introduction}           
\label{sect:intro}
A contact binary is a binary star whose the two components are so close to each other and they merged to share their gaseous envelopes
and they are gravitational bounded to each other \cite{duerbeck1984constraints, muller1903new}. 
There are special sorts of contact binary stars, which are the most ample contact binary stars are W Ursae Majoris (W UMa) contact binaries \cite{mochnacki1981contact}. 
The W UMa type contact binaries (CBs) called as low mass binaries and it includes two stars bonded together and orbiting around their barycenter \cite{boyajian2016planet}.
All stars in the W UMa type CBs are depended on the mass transfer between the 
componets of the three subclass (A-, B-, and W-type) of W UMa system via the physical principle of Roche lobe overflow (RLOF).

In the subclass of W UMa type CBs, the components of the two stars share the same Roche lobe surface as a consequence of these systems.
The stars in these systems fills its Roche lobe then gas flows from the outer layers of that star through the inner Lagrangian point that connects 
the two Roche lobe. The Roche lobe allows the mass transfer between components of W UMa type CBs \cite{eggleton1983approximations}.
The structure of stars in these systems have different properties from single star,
because of the mass transfer between the components of these systems.

The physical parameters of W UMa type CBs were studied  
with the evolution and stellar structure of the stars using the initial and final mass ranges of the W UMa systems.
Depending on the improvements in stellar theories and observational data of W UMa type CBs,
there are mass radius relation (MRR), and mass luminosity relation (MLR) which has been studied by 
\cite{demircan1991stellar, cester1983empirical, gimenez1985mass, yildiz2014origin, sun2020physical}.
But in this study we focused on the investigation of those relations and other relations such as luminosity temperature relation (LTR), radius temperature relation (RTR), and 
mass temperature relation (MTR) using the theoretical zero-age main-sequence (ZAMS) and terminal-age main-sequence (TAMS) models with empirical relations.

Here, in this study we used the catalog of \cite{yildiz2014origin} and the late-type W UMa CBs catalog of \cite{sun2020physical}.
The catalog of  \cite{yildiz2014origin} has different subclass of W UMa type CBs such as A-type and W-type system \cite{binnendijk1970orbital, van1982evolutionary}
whereas the catalog of \cite{sun2020physical} has different subclasses of W UMa type CBs such as A-, B-, and W-type systems \cite{lucy1979observational}.
In these two catalogs, the primary components of A-type are hotter than secondaries, so A-type binary is hotter star 
and their orbital period which is in between 0.4 and 0.8 days \cite{binnendijk1970orbital} whereas the 
primary component of W-type is cooler than the secondaries component and 
it has shorter orbital period in between 0.22 and 0.4 days \cite{binnendijk1970orbital}.
The spectral type of  A- and W-type W UMa type CBs are A-F and G-K \cite{webbink2003contact, rucinski2004contact}, respectively.
Moreover, in the catalog of \cite{sun2020physical} the B-type is poor thermal contact, although it has different effective 
temperatures of the components which are significant \cite{lucy1979observational, sun2020physical}. 
The B-type of late-type W UMa CBs comes from the similarity of their light curves \cite{lucy1979observational}.

\newpage
In this study, we computed the observed and calculated parameters of 100 and 2314 stars from the catalog of \cite{yildiz2014origin} and \cite{sun2020physical}, 
respectively. In these two catalogs, we have studied orbital and stellar parameters to examine the stability of mass transfer and the evolution 
of the stars.

In the present work, we have examined the orbital and stellar parameters of the three subclass of W UMa type CBs using the initial and final mass range 
of these systems. 
Hence, we computed angular momentum loss (AML), critical mass ratio, Roche lobe radius of the donor star, mass ratio, semimajor axis for 
the accretor and donor stars, radius, luminosity and effective temperature, age of the stars, combined and color temperature, using 
the range of the initial and final masses between $ 0.079$ and $2.79 M_{\odot}$ \cite{yildiz2014origin} ; \cite{sun2020physical}.
Also, we investigated the internal stellar structure of the W UMa type CBs using the polytropic model
of the system which study the internal part of the donor stars by introduced stellar structure models in W UMa type CBs \cite{chandrasekhar1957introduction}.

The aim of this study is to make statistical analysis of orbital and stellar parameters of the W UMa type CBs by using the 
catalog of \cite{yildiz2014origin, sun2020physical}. 
Also, we examined the internal stellar structure of W UMa type CBs using the polytropic model.
In these systems, we determined the evolution and stability of mass transfer by applying the ZAMS and TAMS models.
 
This paper is organized as follows: In section 2 we describe the orbital and stellar parameters of W UMa systems.
Section 3 we made a statistical analysis of W UMa type CBs parameter distributions, 
construct distributions of the stellar and orbital parameters of the systems 
between the theoretical ZAMS and TAMS models, and compared them with the empirical relations. In Section 4, we presented the results and
also we parameterized the parameters of those selected systems from the catalog of W UMa type CBs and then compare the results with the
theoretical relations. The conclusions are drowning in Section 5.

\section{Orbital and stellar parameters of W UMa type CBs}\label{sec2}
\subsection{Basic Equations of orbital parameters}
Variation of the orbital parameter with loss of mass from stars were investigated. 
We consider the two accretor mass ($M_{a}$) and donor mass ($ M_{d}$) orbiting each other under the force of gravity.
The orbital angular momentum of the W UMa system is given by \cite{soberman1997stability, siess2011binstar, eggen1961period}; 
\begin{equation}
 J_{orb} =\left(\frac{q \sqrt{G M^{3} a (1-e^{2})}}{(1+q)^{2}}\right),                              \label{eqn:1}   
\end{equation}
where G,  $q=\frac{M_{d}}{M_{a}}$,  a, M = $M_{a} + M_{d}$ , and $e$ are universal gravitational constant, mass ratio, semimajor-axis, 
total mass, and eccentricity of the system, respectively. In this study, we consider that the orbit of the system is circular with $ e=0$.\\
A system should satisfy the third Kepler law, to express the orbital period and semimajor axis of the W UMa type CBs. 
Then, the orbital period of the W UMa system given by 
\begin{equation}
 P_{orb} = \left(\frac{4\pi^{2}a^{3}}{G M}\right)^{\frac{1}{2}}.    \label{eqn:2}
\end{equation}

\newpage
\subsubsection{Roche lobe model}
The Roche lobe is a region where the material is bound to the star by gravity.
The equivalent radius of the Roche lobe overflow of W UMa type CBs is defined based on the radius 
of a sphere with the same volume and the range of mass ratios and the radius of the Roche lobe which is given by \cite{eggleton1983approximations}:
\begin{equation}
 \frac{R_{Ld}}{a} = \frac{0.49 q^{\frac{2}{3}}}{0.6 q^{\frac{2}{3}} + \ln(1 + q^{\frac{1}{3}})},  \label{eqn:3}
\end{equation}
where $R_{Ld}$ is the Roche lobe radius of the donor star.\\
It is difficult to distinguish between the two stars for the purpose of calculating the masses and radii. 
A simpler approximation of Roche lobe radius of the donor star in the Eq. (\ref{eqn:3}) which work well for $0.1 < q < 0.8$
is stated by \cite{paczynski1971evolutionary}.
\begin{equation}
\frac{R_{Ld}}{a} = 0.462 \left(\frac{M_{d}}{M}\right)^{\frac{1}{3}}.         \label{eqn:4}
\end{equation}

\subsubsection{Age of the W UMa type CBs}
The basic thing to get age of the stars is the process of mass transfer between the two components of the W UMa type CBs.
In the W UMa system the mass being transfer from more massive accretor stars to less massive donor stars, and then 
the age of the W UMa type CBs will be changed \cite{bilir2005kinematics}.
The star's age cannot be measured, it only estimates based on the approximation of MS relation
\cite{soderblom2010ages, van2016orbital,negu2018statistical}.
\begin{equation}
\tau_{\odot}=10^{10}\left(\frac{M_{\odot}}{M_{i}}\right)^{2.9}.    ~~~~~~~~~ i\in(a,d)  \label{eqn:5}.
\end{equation}

\subsubsection{Critical mass ratio of W UMa type CBs}
In the W UMa system the critical mass ratio mainly depends on the mass transfer, angular momentum loss
and stellar wind of the W UMa type CBs \cite{chen2008mass, arbutina2009possible}.
The critical mass ratio for W UMa type CBs to predicted as
\begin{equation}
q_{cr}=\frac{(0.362+1)}{{3}(1-\frac{M_{c}}{M})},                 \label{eqn:6}
\end{equation}
where $M_{c}$ is core mass of the contact binary star as noted by \cite{chen2008mass},
and we used the core mass $M_{c}=0.35 M_{\odot}$ for the contact binaries.

\subsection{Stellar parameters of the W UMa type CBs}
The value of each mass of these systems strongly depends on the evolutionary phase of a mass losing star at 
the beginning of the mass loss transfer and the final stage process of A-, B- and W-type of W UMa type CBs.\\ 
\textbf{Case 1}:- \textbf{MRR} and  \textbf{MLR}\\
To determine the radius of the W UMa type CBs, we used the initial and final mass from the two catalogs.
The radius of the star was increased or decreased based on the mass transfer between the components of these systems \cite{gimenez1985mass}.
\newpage
Thus, we investigated the radius based on the mass of A-, W-type and B-type W UMa type CBs.
We illustrated the result of the initial and final luminosity for A-, B- and W-type of the W UMa CBs \cite{cester1983empirical}
by calculating the luminosity of the stars and we compared them with luminosity of the sun ($L_{\odot} = 3.846\times10^{26}W$)
\cite{demircan1991stellar, harmanec1988stellar}.

According to the MRR and MLR, we used for the luminosity, ${L}=0.35\left({M_{i}}\right)^{y}$ where y=2.62 and $M_{i} < 0.7{M_{\odot}}$ 
for terminal-age main sequence star and $L=1.02\left({M_{i}}\right)^{x}$ using x=3.92 and $M_{i} \geq 0.7{M_{\odot}}$ for zero-age main sequence star.

\begin{equation}
\frac{L_{i}}{L_{\odot}}=0.35\left(\frac{M_{i}}{M_{\odot}}\right)^{2.62}, \frac{L_{i}}{L_{\odot}}=1.02\left(\frac{M_{i}}{M_{\odot}}\right)^{3.92}.     ~~~~~~~~~~   \label{eqn:7}
\end{equation}
Although for radius of the stars $R=1.06\left({M_{i}}\right)^{k}$ where k=0.945 and $M_{i} < 1.66{M_{\odot}}$ for TAMS and
$R=1.33\left({M_{i}}\right)^{z}$ we used z=0.555 and $M_{i} \geq 1.66 {M_{\odot}}$ for ZAMS.\\
\begin{equation}
  \frac{R_{i}}{R_{\odot}}=1.06\left(\frac{M_{i}}{M_{\odot}}\right)^{0.945}, \frac{R_{i}}{R_{\odot}}=1.33\left(\frac{M_{i}}{M_{\odot}}\right)^{0.555}.   ~~~~~~          \label{eqn:8}
\end{equation}

\textbf{Case 2}:- \textbf{LTR and RTR}\\
If the stars have low mass, the luminosity will be decreased consequently the temperature also affected 
\cite{demircan1991stellar, cester1983empirical, gimenez1985mass}.
To get the effective temperature of the W UMa type CBs we used the Stefan Boltzmann laws for the luminosity of the stars.
\begin{equation}
 L= 4 \pi R^{2} \sigma T_{eff}^{4}, ~~~~~~ T= \left(\frac{L}{4 \pi R^{2} \sigma}\right)^{0.25}.~~~~~~~        \label{eqn:9}
\end{equation}
Where;- L - luminosity, surface area = $4 \pi R^{2}$, $T_{eff}$ - effective temperature and
Boltzmann constant $(\sigma)= 5.67\times10^{-8}\frac{W}{m^2k^4}$.\\

\subsection{Combined and color temperature}
In the previous studies \cite{sun2020physical} the authors were stated about the combined and color temperature, 
but they didn't calculate these parameters they only give direction how to calculate the two parameters.
The combined and color temperature are used to examine the subclass of W UMa type CBs and their spectral type with period and luminosity relation.
\begin{equation}
T_{c}=\left(\frac{L_{a}+L_{d}}{\frac{L_{a}}{T_{eff_{a}}^{4}} + \frac{L_{d}}{T_{eff_{d}}^{4}}}\right)^{0.25}.            \label{eqn:10}
\end{equation}
where, $T_{c}$ combined temperature, $L_{a},L_{d}$ accretor and donor luminosity and $T_{eff_{a}},T_{eff_{d}}$ accretor and donor effective temperature of the 
W UMa type CBs.\\
Then we expressed the color temperature ($T_{color}$) of these systems \cite{sun2020physical} as $\alpha=\left(\frac{T_{eff_{a}}}{T_{eff_{d}}}\right)^{18.3}$,
which is given by
\begin{equation}
T_{color}=\frac{T_{c}}{\alpha}.             \label{eqn:11}
\end{equation}
\newpage
\subsection{Stability of mass transfer in W UMa type CBs}

The stability of the mass transfer depends on the donor mass loss and donor radius of the donor star and how the conservative process is, 
and also depends on the angular momentum loss processes as well as Roche lobe of the donor star of the W UMa type CBs.
The mass transfer to be continued, the donor must be remained within the Roche lobe and then the 
W UMa type CBs are stable \cite{hjellming1987thresholds, webbink1985stellar}.
The two things are happened at the same time during the stability of mass transfer
in W UMa type CBs:
\begin{enumerate}
 \item a mass-losing star change its radius and
 \item The angular momentum loss could be transferred between the component of the subclass of W UMa type CBs via the donor response: 
\end{enumerate}

\begin{equation}
 \zeta_{L_{d}} = \frac{d\log R_{L_{d}}}{d\log M_{d}}~~~~~ and~~~~~\zeta_{don} = \frac{d\log R_{d}}{d\log M_{d}} ,         \label{eqn:12}
\end{equation}

where $\zeta_{L_{d}}$, is the Roche lobe radius exponents and $\zeta_{don}$ the donor mass radius exponent.
In W UMa system $\zeta_{don}\geq \zeta_{L_{d}}$, implies stability and  $\zeta_{don} < \zeta_{L_{d}}$, implies instability  of mass transfer in these systems
$\zeta_{ad}=\zeta_{don}$ \cite{eggleton1983approximations}. 

\subsection{Internal stellar structure equation for W UMa type CBs}
In this section, we investigated the stellar structure equation of the donor stars in
W UMa type CBs which study the internal part of the stars using the polytropic model.
In polytropic model different number of index are used to explain the internal stellar structure of the donor stars and 
the relation between the physical parameters of W UMa type CBs with stable mass transfer \cite{lane1870theoretical, emden1907gaskugeln, sirotkin2009internal}.
The polytropic model of the W UMa type CBs used to investigated how the pressure of the stars changes with density inside as one move through the W UMa system.
Although, we describe the polytropic model as a function of approximation that is used to represent the equation of state 
of the gas inside stars  \cite{chandrasekhar1957introduction}.
The polytropic stars have relation between the pressure and the density of W UMa type CBs.
Using the central density of the W UMa type CBs, $\rho_{c}$, we have the polytropic model, which is given by 
 \begin{equation}
  P_{r}= K\rho(r)^{1+\frac{1}{n}}=K\rho^\gamma,                          \label{eqn:13}
 \end{equation}
where K and n are polytropic constant and polytropic index. 
In a polytropic stellar model, we assume that the relation between the pressure and density is given by, $ P \propto \rho^{\gamma}$, where $\gamma=\frac{n+1}{n}=1+\frac{1}{n}$.  
Using the famous Lane-Emden equation, we determine the internal stellar structure of the W UMa type CBs and
it's dimensionless \cite{lane1870theoretical, emden1907gaskugeln}.
\begin{equation}
\frac{1}{\xi^{2}}\frac{\mathrm{d}}{\mathrm{d \xi}}\left[{\xi^{2}}\frac{\mathrm{d \xi}}{\mathrm{d \xi}}\right] = -\theta^{n} .                    \label{eqn:14}
\end{equation} 
\newpage
The appropriate boundary condition by applied a polytropic model relation using Eq.(\ref{eqn:14}) are at the center of W UMa type CBs. 
\begin{equation}
when~~~ \xi=0 ~~~~~then~~~~ \theta(0)=1 ~~~~~~~~~~~~~~~~~   \theta'(0)=0 ~~~~~~~~~~~~           \label{eqn:15}
\end{equation}
When pressure and density go to zero in Eq. (\ref{eqn:14}), the surface of the polytropic is located at $\xi=\xi_{1}$,
where $\theta(\xi_{1})=0$.
The analytical solution for different polytropiac index n=0, 1, 5.
\begin{equation}
 \theta_{0}(\xi)= 1-\frac{\xi^{2}}{6}, ~~~~~\theta_{1}(\xi)= \frac{\sin{\xi}}{\xi}, ~~~~~~~~\theta_{5}(\xi)= \left(\frac{3}{3+\xi^{2}}\right)^{\frac{1}{2}}.   \label{eqn:16}
\end{equation} 
\section{Physical parameter distribution of W UMa type CBs}
In this section, we're applying the basic equation of the mass transfer and stellar structure models, 
then we design the theoretical framework for the evolutionary computations of mass transfer in W UMa type CBs.
The analysis of our main sample of the data on W UMa type CBs are from the catalog of \cite{yildiz2014origin} and \cite{sun2020physical} and 
we summarize the details of the observations and calculated parameters in table \ref{tab:1.2} and table \ref{tab:1.3}.
\subsection{Distribution of orbital parameters}
In the evolution of W UMa system, there are different orbital parameter distributions of W UMa type CBs. 

Some orbital parameters for these systems are semimajor axis, orbital period, Roche lobe radius of the donor star and orbital angular momentum of the system.
We determined the orbital parameters by using different initial and final mass ranges and compared their results with the numerical solution of the 
W UMa type CBs in section (4.1).\\ 

\begin{figure}[h!]
\centering
\includegraphics[width=8.50cm, height=6.0cm]{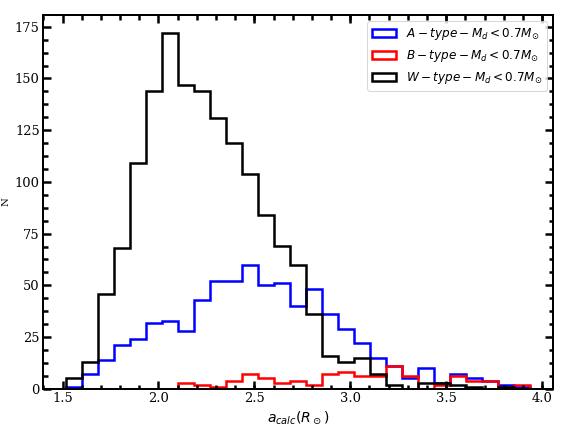}\includegraphics[width=8.50cm, height=6.0cm]{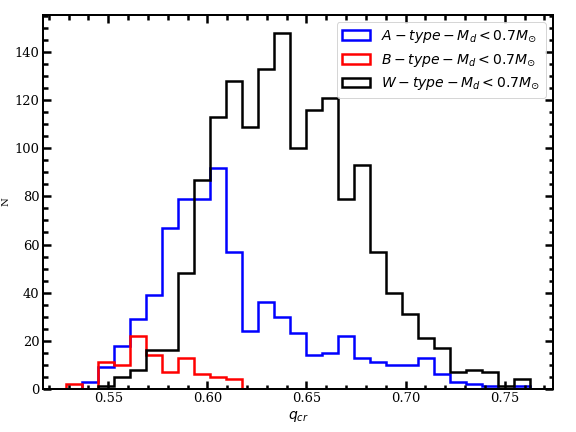}
 \caption{The distribution of calculated semimajor axis with the different ranges of 
 donor mass by applying Eq. (\ref{eqn:2}) and the right side shows distribution of critical mass ratio  
 of A-, B- and W-type late-type W UMa CBs with the donor mass $M_{d} < 0.7M_{\odot}$ using catalog of \cite{sun2020physical}.}
\label{fig:1.1}
\end{figure}
\newpage
The left side of Figure \ref{fig:1.1} shows distribution of the semimajor axis in late-type of W UMa CBs with the different range of final donor mass.
The result of this plot shows the distribution of semimajor axis with variation of colors lines,
the variation of colors is used to identify the subtype of late-type W UMa type CBs.
In this result, the semimajor axis of the late-type W UMa systems shows the expansion and shrinking of the A-, B-, and W-type of late-type W UMa type CBs.
The solid black line shows the W-type star with the peak mode point is 172.0, and it's between 1.936 $R_{\odot}$ and 2.103 $R_{\odot}$.\\

The right panel of figure \ref{fig:1.1} shows the distribution of critical mass ratio for the final stage of donor mass of late-type of W UMa type CBs.
Based on the mass of the stars from catalog of \cite{sun2020physical} and the core mass of the contact binary stars we calculated the critical mass ratio of the 
late-type W UMa CBs using Eq. (\ref{eqn:6}).
The critical mass ratio is used to determine the stability of the mass transfer in W UMa systems.
The peak mode point for solid black line is 148 which indicate the number of W-type CBs that exit in that point and the $q_{cr}$
is between $0.625$ and $0.6417$.
\subsection{Distribution of stellar parameters} 
In the evolution of W UMa systems there are different stellar parameter distributions of W UMa type CBs. 
Some stellar parameters of these systems are luminosity, radius, effective temperature, combined and color temperature.
\begin{figure}[h!]
\begin{center}
\includegraphics[width=8.50cm, height=6.0cm]{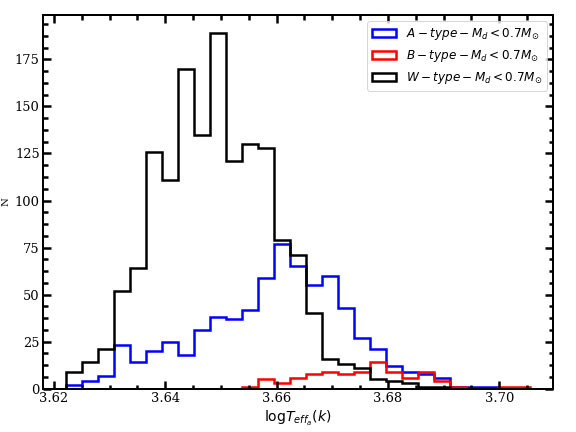}\includegraphics[width=8.50cm, height=6.0cm]{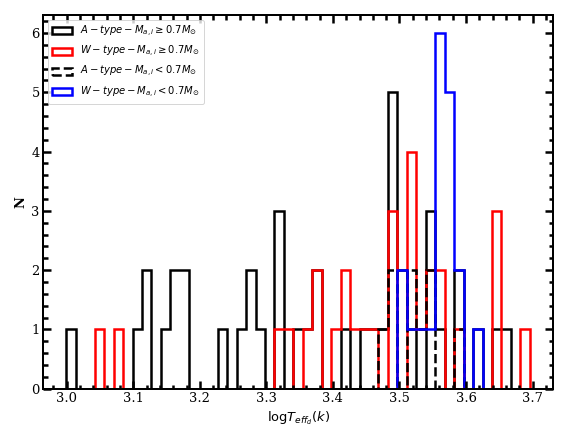}
 \caption{The distribution of the accretor effective temperature with different range of donor masses for the left side figure \ref{fig:1.2}
 and distribution of the donor effective temperature using different range of initial accretor mass of W UMa system based
 on the catalog of \cite{sun2020physical} and \cite{yildiz2014origin} by applying Eq. (\ref{eqn:9}).}
 \label{fig:1.2}
 \end{center}
\end{figure}

The left panel of figure \ref{fig:1.2} shows the distribution of accretor effective temperature using the donor mass range $M_{d} < 0.7M_{\odot}$. 
The A-, B-, and W-type late-type W UMa systems have different effective temperatures in the main sequence regions.
The result the solid black line shows the peak mode point at 189.0 which is in between 3.645 K and 3.650 K for the W-type system.
The solid red line shows the B-type system with high effective temperature and it's a hot star, respectively than the other type of W UMa CBs.
\newpage
The solid blue line shows the A-type star has moderate effective temperature and the solid black line of 
W-type of late-type W UMa CBs has cool effective temperature.
The W UMa type CBs for A- and W-type have identical effective temperature, but there are slightly different between them \cite{yildiz2014origin}. 
The right panel of figure \ref{fig:1.2} shows the variation of effective temperature of the donor stars for A- and W -type of W UMa type CBs.
The black dashed and solid black line distribution shows the effective temperature for A-type of W UMa system and it has
cool effective temperature compared to the W-type of W UMa systems.
The W UMa type CBs for solid blue line of W-type has the peak mode point is 6, which is in between $3.539 K$ and $3.567K$.
\section{Result and Discussion}
\subsection{Numerical solution of W UMa type CBs parameters}
\begin{figure}[htp]
 \centering
 \includegraphics[width=8.50cm, height=6.0cm]{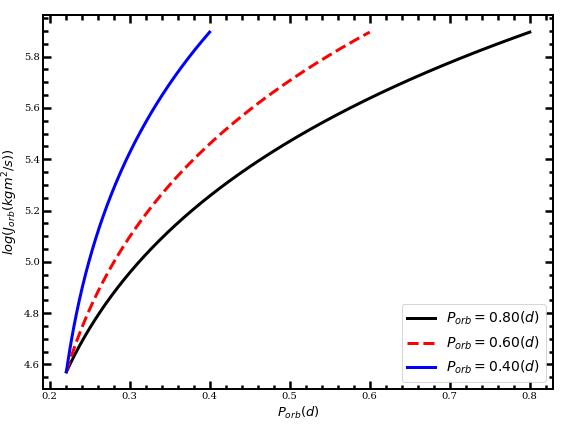}\includegraphics[width=8.50cm, height=6.0cm]{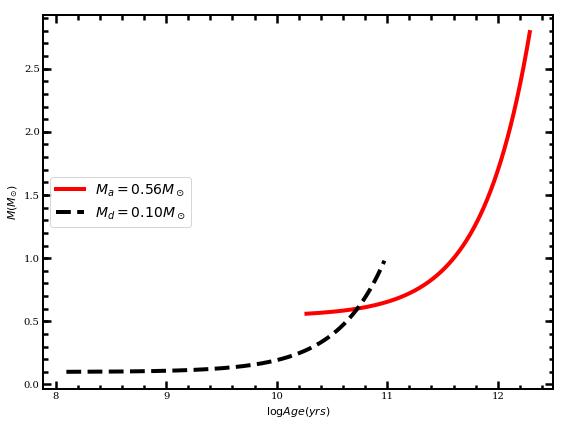}
 \caption{The numerical result of the orbital period and angular momentum loss of the late-type W UMa type CBs
 using the catalog of \cite{sun2020physical} and mass age relation with different range accretor mass of W UMa type CBs using 
 catalog of \cite{yildiz2014origin}.}  
  \label{fig:1.3}
\end{figure}

The orbital angular momentum is used to determine the evolution of W UMa type CBs. 
The rate change of angular momentum loss increases as the orbital separation and orbital period decreases
and then for AML decreases the total mass of the system also decreases \cite{smith2006cataclysmic, patterson1984evolution}.
The left side of figure \ref{fig:1.3} shows the numerical solution of orbital period and orbital angular momentum loss relation 
by applying Eqs. (\ref{eqn:1}) and (\ref{eqn:2}) using different ranges of orbital periods $P_{orb} =0.80, 0.60 $, and 0.40 d.
Angular momentum loss has direct or indirect relation with different orbital parameters, 
although for short orbital period the mass are high and the angular momentum loss also high, so orbital period and AML have indirect relation.
As we observed from the result, the short orbital period of the systems and the AML mechanism were governed by the gravitational radiation 
\cite{patterson1984evolution, taam2001evolution}.\\
\newpage
The right panel of figure \ref{fig:1.3} shows the numerical solution of the final mass distribution and age relation of the W UMa type CBs by applying Eq. (\ref{eqn:5}).
The dashed black line with $M_{d}=0.1 M_{\odot}$ has low mass and it has slow fusion to use their core hydrogen burning, respectively,
it has less pressure, low effective temperature and also low luminosity because these stars have long lifetime in the MS range.
The solid red line shows more massive star with $M_{a}=0.56 M_{\odot}$, so the solid red line shows short lifetime,
have higher central temperatures and high pressures to support themselves against gravitational
collapse and also has fast fusion to used their core of hydrogen burning. Hence, the more massive stars have short lifetime in the MS \cite{soderblom2010ages}.
The A-type shows long lifetime and the W-type also shows short lifetime because of the mass difference
The W-type shows hotter, more massive and more luminous than the A-type W UMa type CBs.\\ 
\subsection{The comparison of initial and final stellar parameters}
\begin{figure}[h!]
\centering
\includegraphics[width=8.50cm, height=6.0cm]{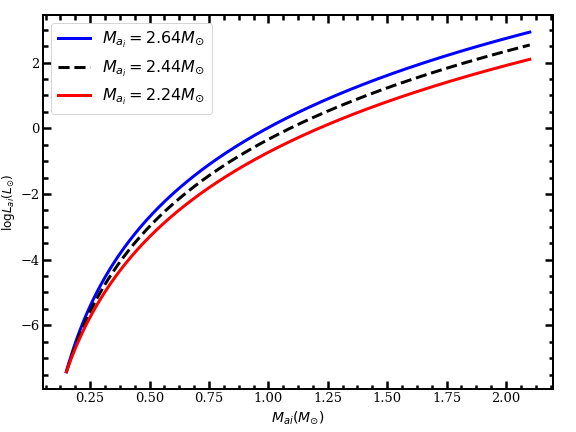}\includegraphics[width=8.50cm, height=6.0cm]{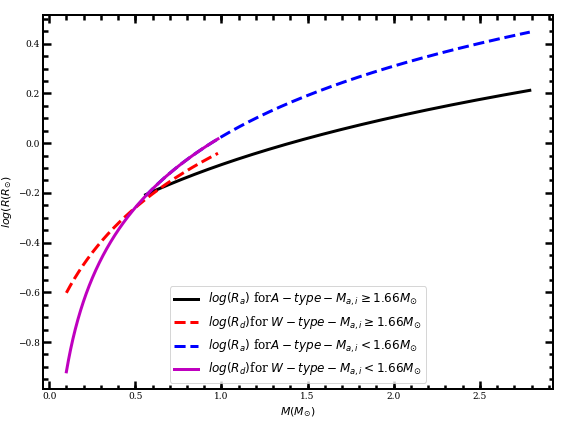}
 \caption{Results between the calculated initial mass luminosity relation and final mass radius relation using 
 the different stellar mass ranges $M_{a,i} \geq 1.66{M_{\odot}}$ and $M_{a,i} < 1.66{M_{\odot}}$
 and by applying Eqs. (\ref{eqn:7}) and (\ref{eqn:8}) for the catalog of \cite{yildiz2014origin}.}
  \label{fig:1.4}
\end{figure}

\textbf{Case 1}:- \textbf{MLR and MRR}\\
The left side figure \ref{fig:1.4} shows the distribution of initial mass luminosity relation of W UMa type CBs depend on the 
theoretical model of ZAMS and TAMS models with empirical relation.
In these relations we incorporated the numerical solution of $\log(L_{a,i})$ and $\log(L_{d,i})$ with
$M_{a,i}$ and $M_{d,i}$, where i is for initial, respectively.
In the ZAMS model we used x=3.92 for $M_{a,i} \geq 0.7{M_{\odot}}$, and for TAMS model we used y=2.62 and $M_{a,i} < 0.7{M_{\odot}}$.
The solid black and red line shows the $\log(L_{a,i})$ and $\log(L_{d,i})$ with x=3.92,
although, the solid blue and the black dashed line shows $\log(L_{a,i})$ and $\log(L_{d,i})$ with y=2.62 for different initial mass ranges.\\
\newpage
The right side of figure \ref{fig:1.4} shows the theoretical result of MRR using different initial accretor mass ranges.
In the W UMa systems the more massive stars have high radius and less massive stars have low radius,
therefore, the radius and the mass of the W UMa stars have direct relation.
The radius of W UMa type CBs are used to examine the phase of stars in the MS region.  
Using Eq. (\ref{eqn:9}) we calculated the initial and the final radius of the W UMa type CBs
with the theoretical models of ZAMS and TAMS.\\

\textbf{Case 2}:- \textbf{LTR and RTR}\\
\begin{figure}[h!]
\centering
\includegraphics[width=8.50cm, height=7.0cm]{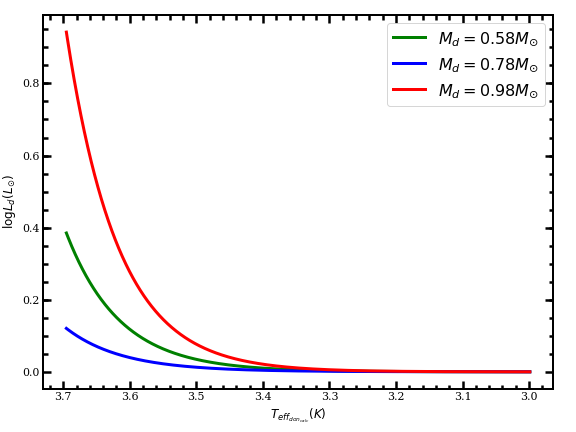}\includegraphics[width=8.50cm, height=7.0cm]{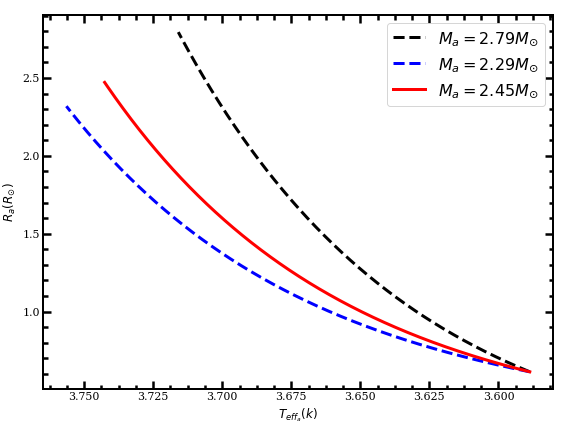}
 \caption{The two numerical results shows the luminosity effective temperature relation, and radius effective temperature relation
 with different range by applying Eq. (\ref{eqn:9}) using catalog of \cite{yildiz2014origin}.}
 \label{fig:1.5}
\end{figure}

The left panel of figure \ref{fig:1.5} shows the numerical solution of distribution final donor luminosity and temperature relation (LTR).
Hence, using different donor mass of W UMa systems $M_{d}=0.58, 0.78$, and 0.98 $M_{\odot}$, 
we obtained different luminosity distribution for W UMa type CBs. 
The solid red line shows  a more massive star with high luminosity as well as high temperature.
The right panel of figure \ref{fig:1.5} shows results of accretor radius effective temperature relation (RTR)
using different range accretor masses, $M_{a}=2.79 M_{\odot}$, $M_{a}=2.29 M_{\odot}$ and  $M_{a}=2.49 M_{\odot}$. 
Although, as we observed from this result when the mass increases the radius also increase, 
but the temperature will be decreases. As we observed from the result, the solid black line has more massive star than the other and it has 
less temperature but has high radius. The solid blue line shows low radius as well as high temperature.\\
 
 \begin{figure}[h!]
\centering
\includegraphics[width=8.50cm, height=7.0cm]{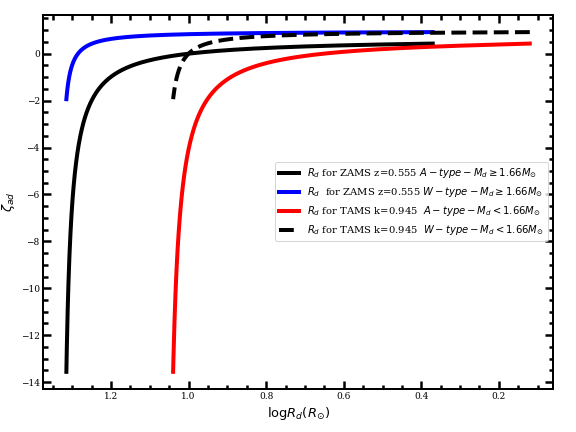}\includegraphics[width=8.50cm, height=7.0cm]{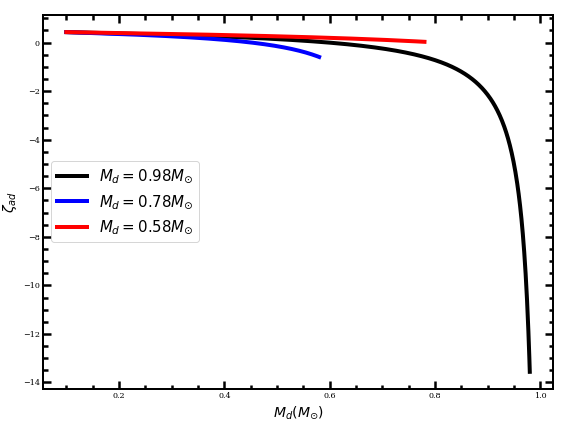}
 \caption{The two numerical result plots shows radius and  donor mass response to the adiabatic ($\zeta_{ad}$), timescale
 of the W UMa type CBs by applying Eq. (\ref{eqn:12}) using catalog of \cite{yildiz2014origin}.}
 \label{fig:1.6}
\end{figure}
The left side of figure \ref{fig:1.6} shows solid black, blue, red lines and dashed black line to represent 
how the radius responds to the adiabatic ($\zeta_{ad}$) calculations using 
theoretical ZAMS and TAMS models with empirical relations by using the donor mass. 
The right side figure \ref{fig:1.6} shows solid black, red, and blue lines to represent how the donor mass responds to the adiabatic ($\zeta_{ad}$) calculations with 
constant mass loss rates at $M_{d}= 0.98, 0.78$ and 0.58 $M_{\odot}$.
\newpage
Comparing both plots the star with lower mass will have long stability period than the star with  greater mass.
In W UMa system, $\zeta_{don}\geq \zeta_{L_{d}}$, implies stability and  $\zeta_{don} < \zeta_{L_{d}}$, implies instability  of mass transfer in these systems
($\zeta_{ad}=\zeta_{don}$) \cite{eggleton1983approximations}. 
These two plots depend on the donor mass of the star because $\zeta_{ad}$, depends on the mass loss for the donor star although
$\zeta_{L_{d}}$, also depends on the angular momentum loss processes for W UMa type CBs.

\begin{figure}[htp]
 \centering
 \includegraphics[width=8.50cm, height=6.0cm]{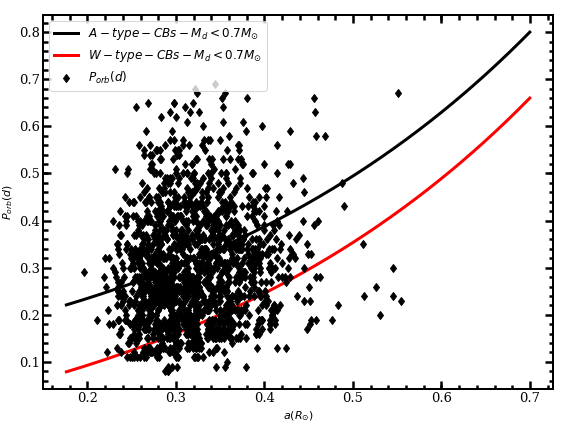}\includegraphics[width=8.50cm, height=6.0cm]{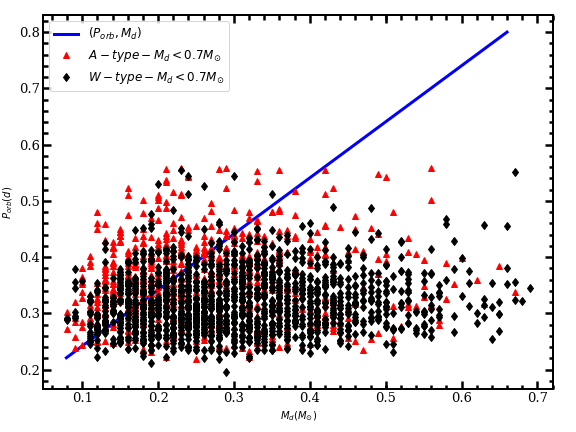}
 \caption{The comparison of orbital period - semimajor axis relation
 and mass-orbital period relation for A-, B- and W-type of W UMa type CBs with donor mass $M_{d} < 0.7M_{\odot}$ by applying Eq. (\ref{eqn:2})  
 \cite{sun2020physical}. }
  \label{fig:1.7}
\end{figure}

The left side of figure \ref{fig:1.7} shows that the solid black and red line shows the relation between 
the semimajor axis and orbital period of A-, W- and B-type CBs.
Hence, we classified the numerical solutions of the semimajor axis and orbital period based on the donor mass $M_{d}< 0.7{M_\odot}$ 
of the late-type W UMa systems \cite{sun2020physical}.
In these results the stars are compacted in one region, this shows the less semimajor axis and short orbital period of the W UMa systems.
Kepler first laws stated about the mass-orbital period relation of the star.
\newpage
As Kepler's stated, the stars are more massive it has short orbital period and when the stars are low massive stars
has long orbital period. Based on the Kepler's laws, the orbital period of the late-type W UMa type CBs are depended on the mass of the stars.
The right side of figure \ref{fig:1.7} shows the final phase distribution of orbital periods-donor mass of the late-type W UMa CBs
compared with A-, B- and W-type CBs.
The A-type of late-type W UMa CBs have long orbital period and W-type has short orbital period \cite{van1982evolutionary}.
The solid blue line shows the numerical solution of these systems.

\begin{figure}[htp]
 \centering
 \includegraphics[width=8.50cm, height=6.0cm]{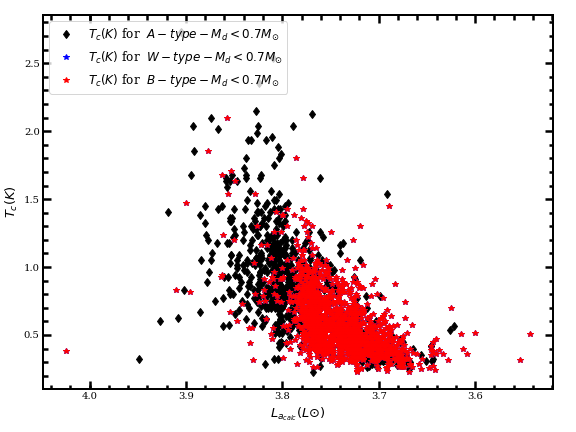}\includegraphics[width=8.50cm, height=6.0cm]{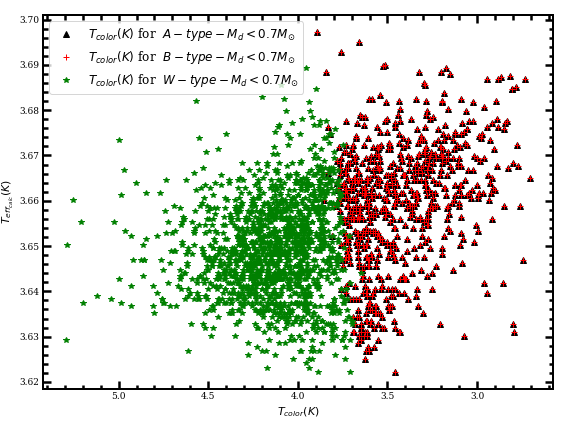}
 \caption{The distribution of combined temperature with the calculated luminosity and color temperature relation, and calculated effective temperature
 relation using different donor mass ranges $M_{d} < 0.7 M_{\odot}$ for A-, B-, and W- type of the late-type W UMa CBs \cite{sun2020physical, pecaut2013intrinsic}
 by applying Eqs. (\ref{eqn:10}) and (\ref{eqn:11}).}
\label{fig:1.8}
\end{figure}

The left side figure \ref{fig:1.8} shows the comparison between the distribution of the combined temperature and the calculated 
effective temperature of the late-type W UMa CBs that can be obtained from the catalog of \cite{sun2020physical}.
In the result the blue-colors for B-type are dominated by the red-color of A-type, this happen
because they have identical combined and effective temperature.\\

The right side figure \ref{fig:1.8} shows the comparison between the color and the effective temperature relation of the late-type W UMa CBs.
Although, using color temperature, we have illustrated either the stars are living in the MS region or it's out of the MS region \cite{pecaut2013intrinsic}.
Stars with high massive, it has fast hydrogen burning, with short life-time, and it has hotter temperature as well as bluer 
\cite{bell2012pre, gullbring1998disk}.
Some of the star has red color, these are because the stars are decreases in the temperature and also having 
slow hydrogen burning in the core, this made the stars to have red color in the main sequence \cite{bell2012pre, gullbring1998disk}.\\
\newpage
\begin{figure}[htp]
 \centering
 \includegraphics[width=8.50cm, height=6.0cm]{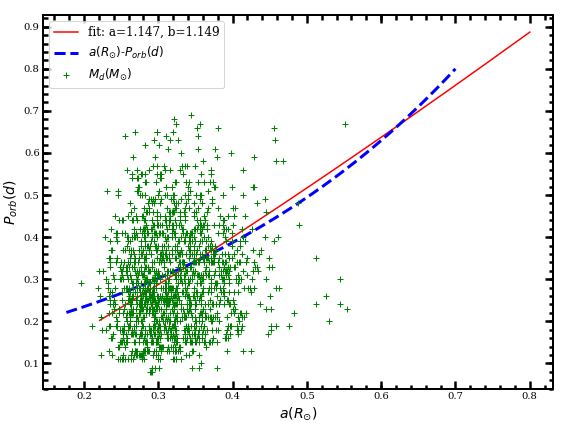}\includegraphics[width=8.50cm, height=6.0cm]{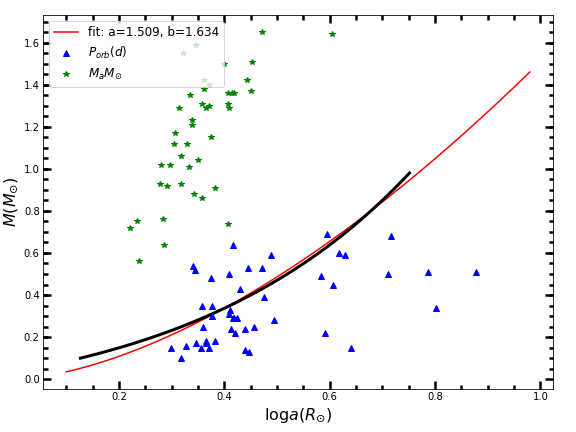}
 \caption{The result of orbital period and semimajor axis relation on the left side panel and the right pane shows the mass 
 and semimajor axis relation that can be obtained from the catalog of \cite{sun2020physical} and \cite{yildiz2014origin}.
 In these results, the distribution shows the fitting function by applying the power law function =$a(x)^b$.}
\label{fig:1.9}
\end{figure}
The left side of figure \ref{fig:1.9} shows the distribution of the orbital period and semimajor axis relation that can be obtained from the catalog
of \cite{sun2020physical} using different parameters of these systems.
In fitting function there are different lows some of them are liner, exponential, power-low, quadratic and others. In this thesis, we used power-law fitting function,
this function used to investigate the error of the catalog of parameters from the catalog of \cite{sun2020physical}.\\
The right side of figure \ref{fig:1.9} shows the fitting function of semimajor axis and mass of the W UMa type CBs using the catalog of \cite{yildiz2014origin}.  
To illustrate the error of given data, we used a power-law fitting function with different fitting parameters "a" and "b".
The solid black line shows the relation between the semimajor axis and the mass W UMa type CBs and the solid red line shows the error function.

\subsubsection{Internal stellar structure of W UMa type CBs}
We explain the internal stellar structure of W UMa type CBs using polytropic model. 
In polytropic model different number of index are used to explain the internal stellar structure and the relation between pressure, radius, density, temperature, 
and mass transfer in the W UMa type CBs with donor star \cite{lane1870theoretical, emden1907gaskugeln, sirotkin2009internal}.
For n=0 the density is constant and the pressure also constant, when n=1 the density of the star keep the radius the same no matters what is mass in the system and also 
radius is independent of mass and the temperature also constant, for n=5 the radius, mass, the central density and central pressure are infinite, 
therefore, the binding energy is infinite.\\
\newpage
The table \ref{tab:1.1}  shows the result  of the polytropic using Lane-Emden equation for each polytropic index,
where $-\theta'_{1}\equiv-\left[\frac{\mathrm{d\theta(\xi)}}{\mathrm{d\xi}}\right]_{\xi=\xi_{1}}$ and $ \xi_{1}=\frac{R}{r_{n}}$.\\

\begin{table}
\centering
\caption{The properties of the Lane-Emden solution using different polytropic index} 
\label{tab:1.1}\
\footnotesize\setlength{\tabcolsep}{3pt}
\begin{tabular}{|l|l|l|l|l|l|l|l|l|l|l|l|}
\cline{3-7}       
\hline
\textbf{$n$} &  0 & 0.5 & 1 & 1.5 & 2 & 2.5 & 3 & 3.5 & 4 & 4.5 &\\
\hline
\textbf{$\frac{R}{r_{n}}$} & 2.449 & 2.753 & 3.142 & 3.654 & 4.353 & 5.355 & 6.897 & 9.536 & 14.972 & 31.837&\\
\hline
\textbf{$-\left[\frac{\mathrm{d\theta(\xi)}}{\mathrm{d\xi}}\right]_{\xi=\xi_{1}}$} & -0.8165 & -0.5000 & -0.3183& -0.2033 & -0.1272 & -0.07626 & -0.04243& -0.02079 & -0.008018 & -0.001715 &\\
\hline                                                                                    
\end{tabular}
\end{table}
\begin{figure}[htp]
 \centering
 \includegraphics[width=8.50cm, height=6.0cm]{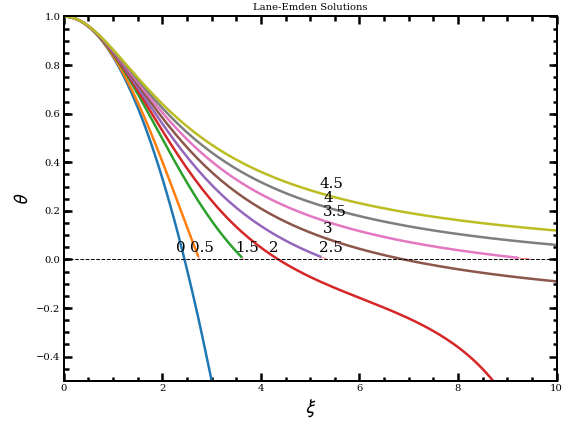}
 \caption{Numerical solution of the internal stellar stracture of the W UMa type CBs for donor star 
 by using Lane-Emden equation Eq. (\ref{eqn:14}) with different polytropic index n = 0, 1, 1.5, 2, 3, 4
 \cite{lane1870theoretical, emden1907gaskugeln}.}
\label{fig:1.10}
\end{figure}
Figure \ref{fig:1.10} shows the numerical solution of Lane-Emden equation by \cite{lane1870theoretical, emden1907gaskugeln}.
The polytropic index is zero the pressure and density are corresponds to constant, and the other analytical and numerical solution 
are existed for n=1, 1.5, 2, 3, 4 and also for 5. Hence, the $ n \geq 5 $ the solution contain infinite mass with the range of $ 0 < n < 5 $ for the polytropic model.
Each color line shows the different polytropic index in figure \ref{fig:1.10}.

\newpage
\begin{table}
\centering
\caption{The observed and calculated parameters for A- and W-type W UMa type CBs using  $M_{d} \geq 0.7 M_{\odot}$ \cite{yildiz2014origin}} 
\label{tab:1.2}\
\footnotesize\setlength{\tabcolsep}{3pt}
\begin{tabular}{l@{\hspace{10pt}} *{11}{c}}
\hline
\cline{3-7}
\textbf{Name}&\textbf{${M_{ai}}$}&\textbf{$M_{di}$}
&\textbf{$L_{a_{i}}$} &\textbf{${L_{d_{i}}}$}&\textbf{$R_{Ld_{i}}$}&\textbf{${R_{ai}}$}&\textbf{${R_{di}}$}&\textbf{$T_{eff_{ai}}$}&\textbf{$T_{eff_{di}}$}&\textbf{OT}&\\
&$(M_{\odot})$ & $(M_{\odot})$& $(L_{\odot})$&$(L_{\odot})$&$(R_{\odot})$&$(R_{\odot})$&$(R_{\odot})$&$(K)$&$(K)$&\\
\hline          
V376 And& 1.89 & 2.50& 12.6 & 37.61 & 1.35& 1.89& 2.21& 3.89& 3.98&A\\ 
NN Vir  & 1.36 & 1.95& 3.40 & 13.98 & 1.47& 1.57& 1.92& 3.79& 3.90&A\\ 
OO Aql  & 0.91 & 1.29& 0.70 & 2.852 & 1.47& 1.26& 1.53& 3.67& 3.78&A\\ 
DN Cam  & 1.5  & 1.86& 4.99 & 11.61 & 1.26& 1.66& 1.87& 3.82& 3.89&W\\ 
EF Boo  & 1.25 & 1.89& 2.44 & 12.62 & 1.57& 1.50& 1.89& 3.77& 3.89&W\\ 
AA UMa  & 1.21 & 1.88& 2.15 & 12.11 & 1.61& 1.47& 1.88& 3.76& 3.89&W\\ 
ER Ori  & 1.26 & 1.74& 2.60 & 8.944 & 1.40& 1.51& 1.80& 3.77& 3.87&W\\ 
SW Lac  & 0.61 & 1.88& 0.14 & 12.11 & 3.57& 1.01& 1.88& 3.55& 3.89&W\\                                                                               
\hline       
\end{tabular}
\end{table} 

\begin{table}
\centering
\caption{The observed and calculated parameters for A-,B and W-type late-type W UMa CBs using  $M_{d} \geq 0.7 M_{\odot}$ \cite{sun2020physical}.} 
\label{tab:1.3}\
\footnotesize\setlength{\tabcolsep}{3pt}
\begin{tabular}{l@{\hspace{10pt}} *{11}{c}}
\hline
\cline{3-7}
\textbf{OT}&\textbf{${M_{a}}$}&\textbf{$M_{d}$}&\textbf{$a_{calc}$}&\textbf{$Age_{1}$}&\textbf{$Age_{2}$}
&\textbf{$q_{cr}$}&\textbf{$T_{c}$}&\textbf{$T_{color}$}&\textbf{$R_{Ld}$}&\\
&$(M_{\odot})$&$(M_{\odot})$&$(R_{\odot})$&$G_{yr}$&$G_{yr}$&&$K$&$K$&\\
\hline
W &1.16& 0.69 &2.57&1.57E+10&3.55E+09&0.36& 3.72& 3.81 &2.50&\\
W &1.16& 0.75 &2.61&1.57E+10&4.51E+09&0.37& 3.73& 3.73 &2.67&\\
W &1.63& 0.98 &3.91&4.19E+10&9.43E+09&0.39& 3.77& 3.79 &3.78&\\
W &1.63& 1.16 &4.01&4.19E+10&1.57E+10&0.39& 3.73& 4.04 &4.38&\\
A &1.12& 0.77 &2.41&1.38E+10&4.86E+09&0.37& 3.72& 3.63 &2.59&\\
B &1.99& 0.89 &3.60&7.35E+10&7.36E+09&0.39& 3.89& 1.74 &2.94&\\
W &1.11& 0.72 &2.45&1.35E+10&3.85E+09&0.36& 3.70& 3.77 &2.50&\\
A &1.40& 0.76 &2.83&2.70E+10&4.68E+09&0.38& 3.76& 3.65 &2.59&\\
W &1.98& 1.32 &4.16&7.24E+10&2.23E+10&0.40& 3.64& 3.98 &4.34&\\
W &0.95& 0.70 &2.19&8.88E+09&3.70E+09&0.35& 3.77& 4.09 &2.46&\\
W &1.21& 0.83 &2.76&1.73E+10&6.03E+09&0.37& 3.74& 3.87 &2.96&\\
W &1.22& 0.72 &2.60&1.78E+10&3.85E+09&0.37& 3.74& 4.03 &2.49&\\
W &1.32& 0.69 &3.05&2.23E+10&3.55E+09&0.37& 3.73& 4.65 &2.74&\\
W &1.22& 0.69 &2.54&1.78E+10&3.55E+09&0.37& 3.76& 4.06 &2.39&\\
A &1.46& 0.72 &2.99&2.99E+10&3.85E+09&0.38& 3.73& 3.61 &2.57&\\
W &1.21& 0.70 &2.56&1.73E+10&3.70E+09&0.37& 3.73& 4.07 &2.45&\\
W &1.17& 0.87 &2.66&1.61E+10&6.67E+09&0.37& 3.75& 3.85 &2.98&\\
A &1.13& 0.74 &2.36&1.46E+10&4.17E+09&0.36& 3.71& 3.60 &2.42&\\
W &1.07& 0.75 &2.31&1.21E+10&4.51E+09&0.36& 3.68& 4.10 &2.51&\\
W &1.28& 0.82 &2.68&2.09E+10&5.82E+09&0.37& 3.73& 4.19 &2.73&\\
W &1.47& 0.81 &3.13&3.05E+10&5.62E+09&0.38& 3.75& 4.11 &2.90&\\
\hline       
\end{tabular}
\end{table}

\newpage
\section{Conclusions}
The reason for the W UMa type CBs pass long-term evolution is because the W UMa type contact binaries known as low mass star.
We have derived the basic evolution equations of mass transfer between the components of W UMa type CBs and we calculated
the orbital and stellar parameters of these selected systems from the W UMa type CBs.\\

We have presented the evolution of angular momentum loss and effect of mass transfer in the subclasses of 
W UMa type CBs based on the catalogs of  \cite{yildiz2014origin} and \cite{sun2020physical}.
We determined the stability of stars based on the effect of mass transfer in the parameters of 
Roche lobe radius of the donor stars and critical mass ratio of W UMa systems.

In this study, we have calculated some stellar and orbital parameters, which are not 
yet calculated in the catalog of  \cite{yildiz2014origin} and \cite{sun2020physical}, some of them were critical mass ratio, 
Roche lobe radius of the donor star, mass ratio, angular momentum loss, orbital angular velocity of the stars of the W UMa type CBs. 
We calculated the combined and color temperature based on the late-type W UMa CBs catalog of \cite{sun2020physical}. 
These two parameters used to explain the color of stars based on the temperature whether they are bluer, red, yellow, orange, and other colors.

We explain the internal stellar structure of W UMa type CBs using polytropic model. 
In polytropic model different number of index are used to explain the internal stellar structure of the donor stars and 
the relation between the physical parmeteres of W UMa type CBs with stable mass transfer \cite{lane1870theoretical, emden1907gaskugeln, sirotkin2009internal}.
The final result are represented in the table \ref{tab:1.2} and table \ref{tab:1.3}.

\begin{acknowledgements}
We thank Space Science and Giospacial Institute (SSGI)-Entoto Observatory and Research Center (EORC)
Astronomy and Astrophysics Department
for supporting this research. This research has made use of NASA’s Astrophysical Data System.
\end{acknowledgements}

 \bibliographystyle{raa}

\label{lastpage}



\begin{thebibliography}{99}
\bibitem[Muller  \& Kempf 1903]{muller1903new}
Muller, G. and Kempf, P. (1903). Blue stragglers. Publications of the Astronomical Society of the Pacific,
105(692):1081.

\bibitem[Duerbeck 1984]{duerbeck1984constraints}
Duerbeck, H. W. (1984). Constraints for cataclysmic binary evolution as derived from space dis-
tributions. In International Astronomical Union Colloquium, volume 80, pages 363–385.
Cambridge University Press.

\bibitem[Patterson. 1984]{patterson1984evolution}
Patterson, J. (1984). The evolution of cataclysmic and low-mass x-ray binaries. The Astrophys-
ical Journal Supplement Series, 54:443–493.

\bibitem[Mochnacki 1981]{mochnacki1981contact}
Mochnacki, S. W. (1981). Contact binary stars. The Astrophysical Journal, 245:650–670.

\bibitem[Eggleton 1983]{eggleton1983approximations}
Eggleton, P.P., 1983. Approximations to the radii of roche lobes. The Astrophysical Journal 268, 368.

\bibitem[Boyajian 2016]{boyajian2016planet}
Boyajian, T. S., LaCourse, D., Rappaport, S., Fabrycky, D., Fischer, D., Gandolfi, D., Kennedy,
G., Korhonen, H., Liu, M., Moor, A., et al. (2016). Planet hunters ix. kic 8462852–where’s
the flux? Monthly Notices of the Royal Astronomical Society, 457(4):3988–4004.

\bibitem[Bell et al. 2012]{bell2012pre}
Bell, C. P., Naylor, T., Mayne, N., Jeffries, R., and Littlefair, S. (2012). Pre-main-sequence
isochrones–i. the pleiades benchmark. Monthly Notices of the Royal Astronomical Society,
424(4):3178–3191.

\bibitem[Gullbring et al. 2012]{gullbring1998disk}
Gullbring, E., Hartmann, L., Briceno, C., and Calvet, N. (1998). Disk accretion rates for t tauri
stars. The Astrophysical Journal, 492(1):323..
\newpage
\bibitem[Yildiz. 2014]{yildiz2014origin}
Yıldız, M. (2014). Origin of w uma-type contact binaries–age and orbital evolution. Monthly
Notices of the Royal Astronomical Society, 437(1):185–194.

\bibitem[Van Hamme. 1982]{van1982evolutionary}
Van Hamme, W. (1982). On the evolutionary state of the W Ursae Majoris contact binaries.
Astronomy and Astrophysics, 105:389–394.

\bibitem[Taam \& Spruit 2001]{taam2001evolution}
Taam, R. E. and Spruit, H. (2001). The evolution of cataclysmic variable binary systems with
circumbinary disks. The Astrophysical Journal, 561(1):329.

\bibitem[Sun et al. 2020]{sun2020physical}
Sun, W., Chen, X., Deng, L., and de Grijs, R. (2020). Physical parameters of late-type contact
binaries in the northern catalina sky survey. The Astrophysical Journal Supplement Series,
247(2):50.

\bibitem[Chen \& Han. 2008]{chen2008mass}
Chen, X. and Han, Z. (2008). Mass transfer from a giant star to a main-sequence companion
and its contribution to long-orbital-period blue stragglers. Monthly Notices of the Royal
Astronomical Society, 387(4):1416–1430.

\bibitem[Lucy and Wilson. 1979]{lucy1979observational}
Lucy, L. and Wilson, R. (1979). Observational tests of theories of contact binaries. The Astro-
physical Journal, 231:502–513

\bibitem[Chandrasekhar and Chandrasekhar 1957]{chandrasekhar1957introduction}
Chandrasekhar, S. and Chandrasekhar, S. (1957). An introduction to the study of stellar struc-
ture, volume 2. Courier Corporation.

\bibitem[Soberman et al. 1997]{soberman1997stability}
Soberman, G., Phinney, E., and Van Den Heuvel, E. (1997). Stability criteria for mass transfer
in binary stellar evolution. arXiv preprint astro-ph/9703016.

\bibitem[Siess et al. 2011]{siess2011binstar} 
Siess, L., Izzard, R., Davis, P., and Deschamps, R. (2011). Binstar: A new binary stellar-
evolution code. In Evolution of Compact Binaries, volume 447, page 339.

\bibitem[Sirotkin \& Kim 2009]{sirotkin2009internal}
Sirotkin, F. V. and Kim, W.-T. (2009). Internal structure and apsidal motions of polytropic stars in close
binaries. The Astrophysical Journal, 698(1):715.

\bibitem[Smith. 2006]{smith2006cataclysmic}
Smith, R. C. (2006). Cataclysmic variables. Contemporary physics, 47(6):363–386.

\bibitem[Eggen 1961]{eggen1961period}
Eggen, O. J. (1961). The period-colour relation for contact binaries. Royal Greenwich Obser-
vatory Bulletins, 31:101–117.

\bibitem[Paczynski. 1971]{paczynski1971evolutionary}
Paczynski, B. (1971). Evolutionary processes in close binary systems. Annual Review of
Astronomy and Astrophysics, 9(1):183–208.

\bibitem[Van et al. 2016]{van2016orbital}
Van Eylen, V., Winn, J. N., and Albrecht, S. (2016). Orbital circularization of hot and cool
kepler eclipsing binaries. The Astrophysical Journal, 824(1):15.

\bibitem[Bilir et al.2005]{bilir2005kinematics}
Bilir, S., Karataş, Y., Demircan, O., and Eker, Z. (2005). Kinematics of W Ursae Majoris
type binaries and evidence of the two types of formation. Monthly Notices of the Royal
Astronomical Society, 357(2):497–517.

\bibitem[Demircan \& Kahraman 1991]{demircan1991stellar}
Demircan, O. and Kahraman, G. (1991). Stellar mass-luminosity and mass-radius relations.
Astrophysics and Space Science, 181(2):313–322.

\bibitem[Soderblom 2010]{soderblom2010ages}
Soderblom, D. R. (2010). The ages of stars. Annual Review of Astronomy and Astrophysics,
48:581–629.

\bibitem[Pecaut \& Mamajek 2003]{pecaut2013intrinsic}
Pecaut, M. J. and Mamajek, E. E. (2013). Intrinsic colors, temperatures, and bolometric correc-
tions of pre-main-sequence stars. The Astrophysical Journal Supplement Series, 208(1):9.

\bibitem[Arbutina 2009]{arbutina2009possible}
Arbutina, B. (2009). Possible solution to the problem of the extreme mass ratio w uma-type
binaries. Monthly Notices of the Royal Astronomical Society, 394(1):501–509.

\bibitem[Binnendijk. 1970]{binnendijk1970orbital}
Binnendijk, L. (1970). The orbital elements of W Ursae Majoris systems. vistas in Astronomy,
12:217–256.

\bibitem[Negu \& Tessema 2018]{negu2018statistical}
Negu, S. and Tessema, S. (2018). Statistical analysis of algol-type eclipsing binaries with stable
mass transfer. Astronomische Nachrichten, 339(9-10):709–717.


\bibitem[Harmanec. 1988]{harmanec1988stellar}
Harmanec, P. (1988). Stellar masses and radii based on modern binary data. Bulletin of the

\bibitem[Hjellming \& Webbink 1987]{hjellming1987thresholds}
Hjellming, M. S. and Webbink, R. F. (1987). Thresholds for rapid mass transfer in binary
systems. i-polytropic models. The Astrophysical Journal, 318:794–808.

\bibitem[Webbink 1985]{webbink1985stellar}
Webbink, R. (1985). Stellar evolution and binaries. Interacting Binary Stars, page 39.

\bibitem[Rucinski 2004]{rucinski2004contact}
Rucinski, S. M. (2004). Contact binary stars of the w uma-type as distance tracers. New
Astronomy Reviews, 48(9):703–709.

\bibitem[Webbink 2003]{webbink2003contact}
Webbink, R. F. (2003). Contact binaries. arXiv preprint astro-ph/0304420.

\bibitem[Rasio \& Shapiro 1994]{rasio1994hydrodynamics}
Rasio, F. A. and Shapiro, S. L. (1994). Hydrodynamics of binary coalescence. ii. polytropes
with gamma= 5/3. arXiv preprint astro-ph/9406032.

\bibitem[Lane 1870]{lane1870theoretical}
Lane, H. J. (1870). On the theoretical temperature of the sun, under the hypothesis of a gaseous
mass maintaining its volume by its internal heat, and depending on the laws of gases as
known to terrestrial experiment. American Journal of Science, 2(148):57–74.

\bibitem[Emden 1907]{emden1907gaskugeln}
Emden, R. (1907). Gaskugeln: Anwendungen der mechanischen Wärmetheorie auf kosmolo-
gische und meteorologische Probleme. BG Teubner.

\bibitem[Cester et al. 1983]{cester1983empirical}
Cester, B., Ferluga, S., and Boehm, C. (1983). The empirical mass-luminosity relation. Astro-
physics and space science, 96(1):125–140.

\bibitem[Gimenez \& Zamorano 1985]{gimenez1985mass}
Gimenez, A. and Zamorano, J. (1985). The mass-radius relation in binary systems. Astro-
physics and Space science, 114(2):259–269.

\end{thebibliography}

\end{document}